\documentstyle[preprint,prl,aps]{revtex}
\begin{document}
\draft
\title{Inelastic Quantum Transport}
\author{Janez Bon\v ca$^{a,b}$ and S. A. Trugman$^a$}
\address{$^a$Theory Division, 
Los Alamos National Laboratory, 
Los Alamos, NM  87545, 
$^b$J. Stefan Institute and
University of Ljubljana, 1001, Slovenia}
\date{\today}
\maketitle
\begin{abstract}

We solve a Schr\" odinger equation for
inelastic quantum transport that retains
full quantum coherence, in contrast to
previous rate or Boltzmann equation approaches.
The model Hamiltonian
is the zero temperature 1d Holstein model for an electron
coupled to optical phonons (polaron), in a strong electric field.
The Hilbert space grows
exponentially with electron position, forming
a non-standard Bethe lattice.
We calculate nonperturbatively
the transport current, electron-phonon correlations,
and quantum diffusion.  This system
is a toy model for the constantly
branching ``wavefunction of the universe''.
\end{abstract}

This paper is a fundamental study of quantum transport in the presence of
inelastic degrees of freedom.  
We retain coherent quantum effects
in a way that has not to our knowledge been done previously.
There are a number of relevant model
systems, but for concreteness we consider an electron 
interacting with optical phonons (the Holstein model),
driven by a strong electric field.
Our calculation is variational.  It is non-perturbative
in the electron-phonon coupling, the electric field,
and everything else.  

Most previous work on this problem has relied on
rate or Boltzmann equations, which are valid only for weak
electron-phonon coupling and ignore
quantum coherence effects \cite{emin,roy,kummel,wilkins}.
These theories calculate transition rate probabilities, rather
than complex quantum amplitudes.
It is not clear how accurate such a treatment is,
since, for example, a polaron is a quantum coherent object.
This work takes a different approach.  
We calculate the wavefunction $| \psi \rangle$
that is the scattering solution to the
Schr\" odinger equation $ H | \psi \rangle = E | \psi \rangle $.
The solution consists of the complex amplitude and phase for
each basis state in the many-body Hilbert space.
The results are relevant to
high field transport experiments
on semiconductors such as ZnS \cite{maekawa},
tunnel cathode structures \cite{handy,savoye},
and microchannels \cite{tsui}.  They may
also be relevant to systems exhibiting
polaron hopping conductivity, such as
colossal magnetoresistance (CMR) materials \cite{cmr}.
There has been related theoretical work
that takes a very different (path integral)
approach to the Fr\" ohlich polaron problem, which
has long-range electron-phonon interactions and parabolic bands
\cite{thornber,sauls}.  

We consider 
the Holstein model describing a single
electron in a one-dimensional tight-binding lattice,
which interacts locally with dispersionless optical
phonons \cite{emin}. 
(Other models of inelastic quantum transport can be
treated by similar methods.)
A constant electric field drives the electron.
The Hamiltonian is
\begin{eqnarray}
H &=& \sum_j \epsilon _j c^\dagger _j c_j  
-t \sum_j  ( c^\dagger _j c_{j+1} + h.c.) \nonumber \\
&+& \Omega \sum_j  a^\dagger _j a_j 
- \lambda \sum_j  c^\dagger _j c_j  (a^\dagger _j + a_j),
\label{ham}
\end{eqnarray}
where $t$ is the electron hopping, $\Omega$ is the phonon frequency,
and $\lambda$ is the electron-phonon coupling.
We consider only the case of zero temperature in this paper.
Since there is no linear response for this problem
at $T=0$, the electric field $\cal E$ appears explicitly in
the Hamiltonian rather than being added as a perturbation.
The electric field causes the site energy 
$\epsilon _j$ to decrease linearly with position,
$\epsilon _j = -e {\cal E} a j \equiv - \alpha j$,
where $a$ is the lattice constant
and $j$ is the site index.
In contrast to linear response theory, the
explicit electric field in $H$ allows one to consider
the {\it nonlinear} current carried in a large field.
The lattice extends to infinity in both directions, and
is translation invariant except that the diagonal site
energy decreases linearly with distance.

An electron injected at the origin
initially accelerates to the right
in the electric field.  If the electron-phonon
coupling $\lambda$ were zero, the kinetic energy
of the electron would increase until it
reached the top of the local band of energetically allowed states.
The electron would then reverse direction and
undergo Bloch oscillations, with AC but no DC current,
characteristic of
Wannier-Stark levels \cite{kummel}.
For nonzero electron-phonon coupling, the Wannier-Stark
levels decay and the electron
continues to move to the right on average, leaving
a trail of phonons behind it, like a particle
emitting \v Cerenkov radiation.

We seek to solve as completely as possible
the Schr\"odinger equation $H | \psi \rangle = E | \psi \rangle $,
with the Hamiltonian in Eq.~(\ref{ham}).  This is a formidable
problem, because by the time the electron has traveled
downhill $j$ sites, it has emitted on the average
$ \alpha j / \Omega $ phonons.  These phonons can be
distributed in exponentially many ways on the sites
uphill from the electron.  Thus the relevant part of the Hilbert
space for the electron on site $j$ increases exponentially
with $j$.  
This problem can nevertheless be numerically solved
exactly in an (exponentially growing) variational Hilbert space \cite{janez}.

We define a variational Hilbert space, which can be 
systematically enlarged until the solution has converged.
A complete set of basis states 
of the many-body Hilbert space 
can be written
$ |M \rangle = |j; \dots , m_{-1}, m_0, m_1, \dots \rangle$, where $j$
is the electron position and $m_n$ is the number of phonon
quanta on site $n$.  The diagonal energy of the state is
$E_M = - \alpha j + \Omega ( \sum _n m_n )$.
A variational space is defined as follows.
A basis state is retained only if:
(1)  The number of phonon quanta on a given site
does not exceed a constant $n_{max}$.
(2)  The diagonal energy $E_M$ of a many-body basis state
$| M \rangle $ lies in a given energy range,
$\Delta _1 \leq E_M \leq \Delta _2$.
(The accuracy is greatest if the desired energy eigenvalue $E$
is near the center of the range.)
These constraints are not terribly restrictive, 
as $n_{max}$ and the energy range can be systematically increased
to check that the solution has converged.  
For technical reasons that will be explained below,
we also apply a third condition:  
(3)  all phonon excitations may appear only on the same site
or left (uphill) of the electron.  This condition
is motivated by the fact that for a strong electric
field, by analogy with \v Cerenkov radiation,
almost all of the phonons are left behind the electron.
This condition can also be relaxed, but with some effort.

To illustrate the method, we first consider one of the simplest
nontrivial cases, 
with $\alpha = \Omega = \Delta _2 = -\Delta _ 1 = 1$, shown in
Fig. (\ref{bethe}) \cite{largee}.  
The electron is injected into state $|1 \rangle $, which has 
the electron at the origin and no phonon excitations.
It can
either hop to the right (state $|3 \rangle $ ), or create a phonon
(state $|2 \rangle $ ) \cite{left}.  The lattice extends to infinity, with
the number of possibilities
increasing exponentially as the electron moves downhill.
The off-diagonal matrix elements are $-t$ on bonds $a$ and $b$,
$-\lambda$ on bonds $x$ and $y$, and $-\lambda \sqrt {2}$
on bond $z$.  The diagonal energies are $-$1, 0, or +1.
To take a specific example,
the Schr\"odinger equation on site (4) is
\begin{equation}
E \psi _4 = \epsilon _4 \psi _4 -t \psi _2 -t \psi _8 - \lambda \psi _6 ,
\label{schrodinger}
\end{equation}
where the diagonal energy $\epsilon _4 = 0$.
The exponential growth in the number of states
should be contrasted with the standard polaron problem where the electric
field ${\cal E} = 0$, which is shown in Fig. (\ref{noe}).

We seek a scattering solution for the lattice of
Fig. (\ref{bethe}) where the electron is injected on site 1,
and generalized currents flow 
along the bonds.
The Schr\"odinger equation is satisfied on
all sites (other than 1) with an energy eigenvalue $E$
that is known in advance.  The lattice
has no loops, and is in fact an unusual type
of Bethe lattice.  
The entire lattice connected downstream (to the right) of any $b$ bond is the
same as that for any other $b$ bond.  
This leads to the condition that in the 
scattering eigenfunction,
the amplitude ratio across any $b$ bond is a fixed
complex number that will be called $b$,
$~ \psi_3 / \psi_1 = \psi_8 / \psi_4 = \dots \equiv b$.
The same can be said of $x$ bonds, etc.\cite{imped}

The scattering problem 
can be solved by taking a finite piece
of the lattice, as shown in Fig.~(\ref{bethe}).
Consider a terminal site $j$,
whose downstream neighbor $m$ on a $b$ bond has been omitted.
The Schr\"odinger equation on site $j$ is written
assuming $\psi_m / \psi_j = b_0$, where $b_0$
is an initial guess for $b$. 
Proceeding similarly for the other terminal sites,
the system is solved for all amplitudes $\psi _ i$.
A new set of complex numbers $a_1, b_1, \dots$
is obtained by taking ratios of the amplitudes on specified
interior bonds.  This new set of complex numbers is then used on
the edge sites, and the equations iterated until
self-consistency is achieved.  
The initial complex numbers
are chosen with a positive imaginary part,
corresponding to outgoing solutions.  
The solution is obtained numerically by solving a sparse complex
system of equations of the form $A x = b$.  (Since the eigenvalue
$E$ is known in advance, the numerical routine required
is for simultaneous linear equations \cite{y12}, not for eigensystems.)
Once the wavefunction has been obtained \cite{answer}, the expectation
of any observable can be calculated.  
If the maximum number of phonons per site in condition
(1) or the energy range in condition (2) are increased,
the representation of the problem 
still has no loops.
The graph does, however,
contain more inequivalent bonds.  We have solved
this problem with hundreds of inequivalent
bonds (rather than the 5 types discussed above)
by the same iterative method \cite{unique}. 
If condition (3) above is relaxed, and phonons are allowed
to the right of the electron as well, then a low
density of loops appears in the graph. 
The loops lead to complications that
will be discussed below.

It is sometimes asserted
in transport theory that an electron
scattering from a static impurity retains its
phase, but that when a phonon is emitted or
absorbed, the phase is randomized.  We do not
subscribe to this point of view. 
Every many-body basis state in Fig.~(\ref{bethe})
has a wavefunction $\psi _j$ with both an amplitude
and definite phase.
Some observables,
such as the electron density on a particular site,
make no use of the phase information in this basis---they
would be unchanged if all phases were randomized.
Other observables, however, depend crucially
on the phases.

Consider the average displacement of the oscillator
on site $j$ when the electron is on site $m$,
$\langle ( a^\dagger _j + a_j ) c^\dagger _m  c_m \rangle $,
plotted in Fig.~(\ref{correlation}).
This expectation depends crucially on the relative
phases of the different $\psi _i$.  
It is straightforward to show that as
the real physical distance $m-j$ between the electron
and phonon in the correlation function 
increases, the expectation requires knowledge
of the relative phases of many-body basis states
that are farther apart in the Hilbert space, Fig.~(\ref{bethe}).
(The metric distance $l$ between basis states in the many-body
Hilbert space shown in Fig.~(\ref{bethe}) is defined as the minimum
number of bonds to get from one to the other.)
Although distant phase correlations are required
for large $m-j$, 
the sum of the many terms is found to be small.
In fact, for large $m-j$,
one would not obtain a seriously wrong answer by
taking the computed amplitudes and averaging over
all random phases, which would give zero for the correlation function.

Also plotted in Fig.~(\ref{correlation}) is 
the phonon energy on site $j$ when the electron
is on site $m$, $\langle  a^\dagger _j  a_j  c^\dagger _m  c_m \rangle $,
which goes to a constant for large $m-j$
(except near the injection site).
The electron sets a phonon ringing as it passes.
As the electron moves far away, the energy remains
in the oscillator (it has nowhere to go),
but the oscillator is as likely to have a positive as a negative displacement.  
These phonon excitations should be measurable
by neutron scattering.
The correlations plotted
as triangles in Fig.~(\ref{correlation})
cannot be calculated by the methods of Ref. \cite{emin}.
(It has generally been assumed in previous work that the phonons
begin and remain in equilibrium.)

We now consider the current
or average electron velocity.
Figure (\ref{current}) plots the velocity
$ v(E)  = \langle J(E) \rangle / \langle n(E) \rangle $,
the ratio of the particle current to the particle density,
as a function of the total energy eigenvalue $E$.
The solid curves are for different values
of the electron-phonon coupling $\lambda$,
and the dashed curve is for a smaller electric
field $\alpha = 1/2$.
The velocity is a periodic function of the energy $E$
with period $\alpha$,
because a shift in the
energy by $\alpha$ is equivalent to
a shift in the origin by one lattice constant.
The dashed curve
has period 1/2, and is more nearly
constant than the others.
The fact that the numerically calculated $v(E)$ 
is fairly accurately periodic is a confirmation that
the variational energy range $[ \Delta _1 , \Delta _2 ]$
is large enough that the answer has converged
for the energy eigenvalues $E$ plotted.
For small electron-phonon coupling $\lambda$, 
there are energies $E$ (the eigenvalue
of the entire coupled electron-phonon system) where
the electron cannot propagate at all.  This is
analogous to band gaps in crystals.  The forbidden
regions are perhaps less surprising when one
considers that if 
the electron-phonon coupling
$\lambda$ were zero, 
the electron would be localized in a Wannier-Stark state,
and there would be no current at any energy.
We believe that these forbidden regions 
would disappear if the optical phonons were given nonzero dispersion.

It is more difficult to calculate the velocity $v(\alpha)$ as
a function of the electric field, because this is a nonanalytic function.
When $\alpha / \Omega$ is a low order
rational, the representation of
the Hilbert space as shown in Fig.~(\ref{bethe})
contains subsets that are 1d periodic tight-binding
lattices with a small basis, such as the branch $byby \dots .$
In contrast, for $\alpha / \Omega$ irrational, all such subsets
are quasiperiodic, leading to different transport.
A similar argument accounts for the energy dependence of $v$
when $\alpha / \Omega$ is rational:  it matters 
how close the diagonal energies
are to the eigenvalue $E$.
Some aspects of these results differ from
those obtained previously.  Using rate equations
that neglect quantum coherence, Emin and Hart
(and others)
obtain an infinite drift velocity for dispersionless
optical phonons \cite{emin}, in contrast to the finite result
of Fig.~(\ref{current}).  
We are, however, in qualitative agreement with Ref. \cite{emin}
in that reducing the electric field in this regime
increases the current. 
We believe that the enhanced nonlinear conductivity
measured by Maekawa in $\rm ZnS$ occurs
at rational values of $\alpha / \Omega$.
In contrast to his interpretation \cite{maekawa}, this need not
signal the existence of long-lived Wannier-Stark levels;
see also Ref. \cite{kummel}.

We now consider the time evolution of a wavepacket,
formed by injecting an electron
into site (1) with
$\psi _1 = \exp (  -i E_0 t  -t^2 / 2 \tau ^2 )$.
For a packet that is narrow in energy space (wide in real space),
the time evolution depends on the first
and second energy derivates of the amplitude
and phase of the complex numbers $\{ a, b, x, \dots \} $.
A detailed calculation \cite{janez2} shows that
the width increases in time as
\begin{equation}
\langle [ x(t) - \bar x (t) ]^2 \rangle = 
\sigma_0 ^2
+ \mu t  + \nu t^2 .
\label{spreadmanyb}
\end{equation}
The last term is a quantum coherent spreading
similar to the one that appears in a simple 1d chain (without
branches), with $\nu$ decreasing
as the initial wavepacket gets wider,
$\nu \sim \sigma _ 0 ^ {-2}$  \cite{tsquare}.
The additional term is linear in time, with the coefficient
$\mu$ independent of the initial width of the wavepacket.
It describes the same behavior as
the {\it classical } diffusion of particles
due to Brownian motion.  It is perhaps surprising to see such
behavior emerge from a fully coherent quantum mechanical calculation
(without disorder) at zero temperature.
One reason the $\mu t$ term appears in the branching Hilbert space
is that the energy variation of the
modulus of the complex numbers causes parts of the wavepacket
with slightly different energies to move preferentially into different
branches of the tree.  (This property is not present
in standard Bethe lattices \cite{economou}.)
Once on different branches, these
amplitudes cannot destructively interfere, which
causes the wavepacket to spread \cite{interfere}.
Another reason 
is that the group velocities differ on different branches.

Another question
is how much entropy is generated
as the electron travels downhill.
The technical answer is that no entropy is generated.
The system begins in a pure state and remains in a pure
state under time evolution of the Schr\"odinger equation \cite{entropy}.

When phonon excitations are allowed to the right of
the electron (relaxing condition 3), 
the structure of the Hilbert space shown
in Fig.~(\ref{bethe}) acquires
extra states that form infrequent loops.  The structure
is no-longer exactly tree-like, but it is still self-similar.
The problem can still
be solved by similar methods (finding ``images'' of
the terminal sites nearer the origin of the lattice),
and also by novel methods \cite{janez2}.
None of these methods is as yet completely satisfactory, however.
One problem is that the iterative
scheme may fail to converge for certain energies.
We will not discuss this extension in any detail
in this publication.

A final point concerns the foundations of quantum mechanics.
The ``wavefunction of the universe'' has been described
as residing in an abstract space that is constantly branching \cite{everett}.
One question that has been considered is the issue of
why amplitudes on different
branches seem not interfere with each other.  
The usual answer is known as the process of
decoherence \cite{zurek}.
The Holstein polaron in an electric field,
with a Hilbert space illustrated in Fig.~(\ref{bethe}),
provides a simple toy model
for investigating such questions.  As described above, 
there is noticeable interference at least for short times
for the Holstein model.

{\it Acknowledgments:}
We would like to thank Elihu Abrahams, Alan Bishop, 
Sudip Chakravarty, Jan Engelbrecht, Emil Mottola, 
Heinrich R\" oder,
Douglas Scalapino, Robert Schrieffer, 
James Sethna, Richard Silver, Antoinette Taylor, 
John Wilkins, Jan Zaanen,
and Wojciech Zurek for valuable
discussions.
We would also like to acknowledge the Aspen Center for Physics,
where part of this research was done.
This work was supported by the U.S. DOE.

\begin{figure}
\caption[]{ A variational many-body Hilbert space for
an electron coupled to phonons in an electric field.
Each dot represents a basis state
in the many-body Hilbert space.  Lines represent
nonzero off-diagonal matrix elements.  
The many-body problem
can alternatively be viewed as a 1-body
problem, where the dots are Wannier orbitals
with hopping matrix elements shown by the bonds.} 
\label{bethe}
\end{figure} 

\begin{figure}
\caption[]{ A variational many-body Hilbert space
analogous to Fig. (\ref{bethe}) is shown
for the case of zero electric field.
Only states with 0 or 1 phonon excitation on sites
not farther than one lattice spacing from the electron
are included. 
State $|1 \rangle$ has no phonon excitations.
Vertical bonds create phonons, and (nearly) horizontal
bonds are electron hops.
For ${\cal E} = 0$, the
lattice is translation-invariant, does not grow
exponentially, and can be solved simply in $\vec k$-space
\cite{evora} .
}
\label{noe}
\end{figure}

\begin{figure}
\caption[]{ 
The oscillator coordinate $x_j$ when
the electron is on site $m$,
$\langle ( a^\dagger _j + a_j ) c^\dagger _m  c_m \rangle $,
plotted as a function of $j$ for $m=8$ (triangles).
The oscillator
energy $\langle  a^\dagger _j  a_j  c^\dagger _m  c_m \rangle $
as a function of $j$ (squares).
The electron is injected into the system 
on site 0.
Parameters are $\lambda = 0.5$,  $\alpha = \Omega = t = 1$,
$[\Delta _ 1, \Delta _ 2] = [-3,3]$, $E=0.6$, with up to three phonons
per site.
}
\label{correlation}
\end{figure} 

\begin{figure}
\caption[]{ (a) The drift velocity $v$ (current/density)
is plotted as a function of the energy eigenvalue $E$ for 
electron-phonon coupling constant 
$\lambda = 0.3, 0.5,$ and 0.7, from
the bottom curve up.
For the smaller $\lambda 's$, there
are forbidden energies at which there are no propagating states.
Other parameters are
$\alpha = \Omega = t = 1$.
Basis states are retained in the diagonal energy range 
$[\Delta _ 1, \Delta _ 2] = [-4,4]$, with up to two phonons
per site.  
The dashed curve is for electric field
$\alpha = 1/2$, $\lambda = 0.5$.
It has been shifted up by 0.15 .
}  
\label{current}
\end{figure}


\begin{references}
\bibitem{largee}  A field $\alpha = \Omega$
is a very large electric field, corresponding to a potential
drop of an optical phonon frequency per lattice site.
Experiments are, however, done in this 
regime \cite{maekawa,handy,savoye}.
One could alternatively consider a site in the Hamiltonian
to represent a cluster of several atoms, which would
reduce the physical electric field.
\bibitem{left}
The electron could also hop to the left, but this 
dead-end branch does not affect the results,
and is omitted.
\bibitem{imped}  The infinite tree attached downstream
from a given site can be regarded as a lumped impedance,
similar to an antenna problem.
\bibitem{answer} If each site in Fig. (1) is labeled by the sequence
of bonds that are followed to reach it, the question
``what is the wavefunction on site $j$?'' answers itself.
For example, $\psi_{11} = \psi_{byza} = byza$,
the product of the self-consistent complex numbers.
\bibitem{unique} 
A unique solution is always found for outgoing boundary
conditions.  The complex numbers
$\{ a, b, \dots \}$ 
are the same on all equivalent bonds, and
do not change to machine accuracy if the lattice is expanded
to include additional generations.
\bibitem{tsquare}  It is clear from Fig. (4) 
that since the wavepacket is composed of
a range of energies, and each energy has a different
velocity, the width-squared $\langle [ x(t) - \bar x (t) ]^2 \rangle$
must increase as $t^2$ in the long time limit.
\bibitem{interfere}  In the simple 1d chain, one can write
the wavefunction as a sum of plane waves.
Each plane wave has infinite spatial extent;  the sum
has a finite extent because of destructive interference.
\bibitem{entropy}  We have, however, calculated a basis-dependent
``entropy'' $\tilde S$
as the Shannon entropy
in the phonon sequence left behind, assuming that
all phase information has been lost.  
The result is 
$\tilde S = \gamma N$, where $\gamma$ is a constant
that depends on coupling constants.  The calculation
is somewhat complicated by the fact that the probability
of finding $m_j$ phonons on site $j$ depends on $m_{j-1}$.
We emphasize, however, that $\tilde S$ is not a true
entropy and that the value of $\tilde S$ is basis-dependent.
\end{references}
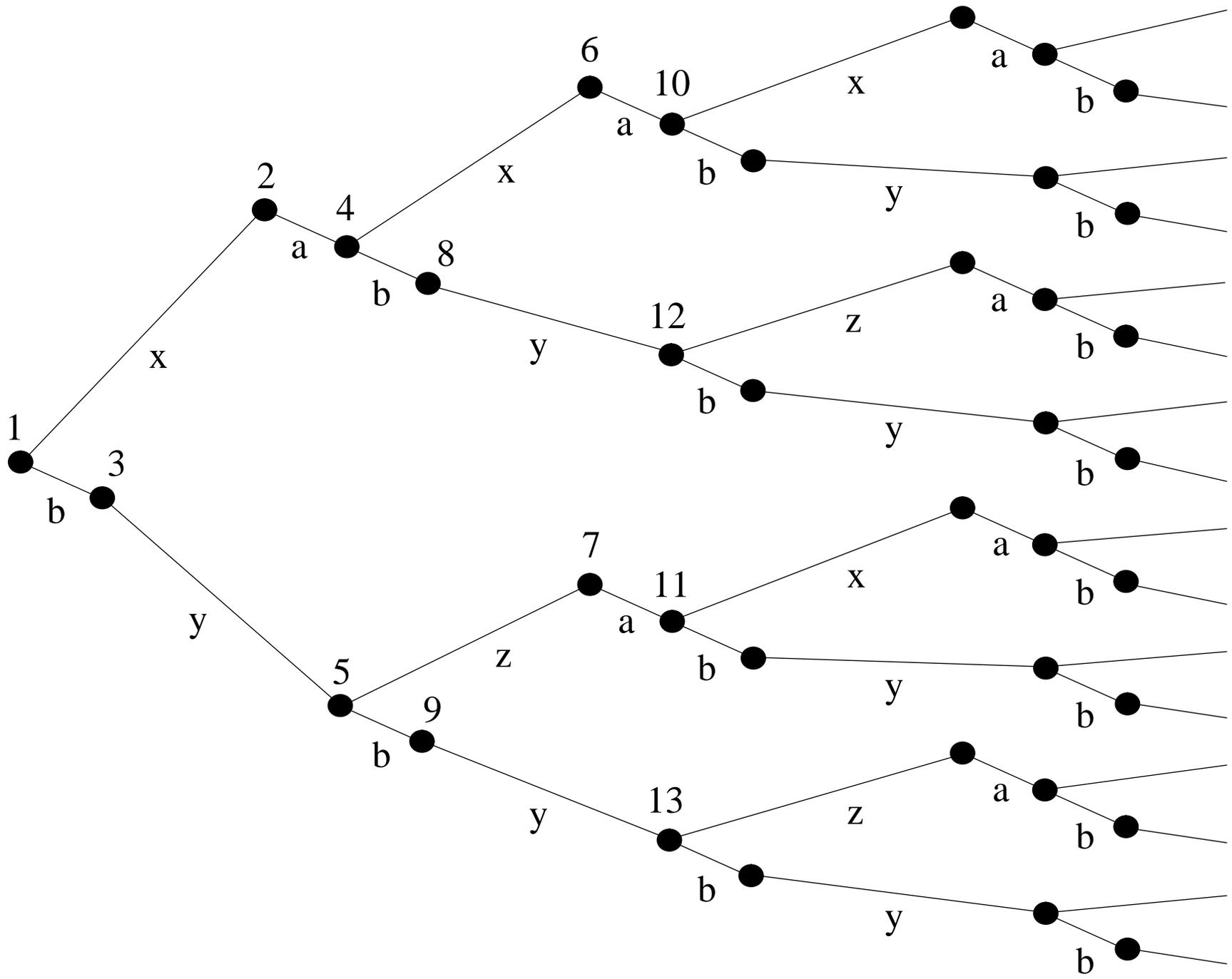
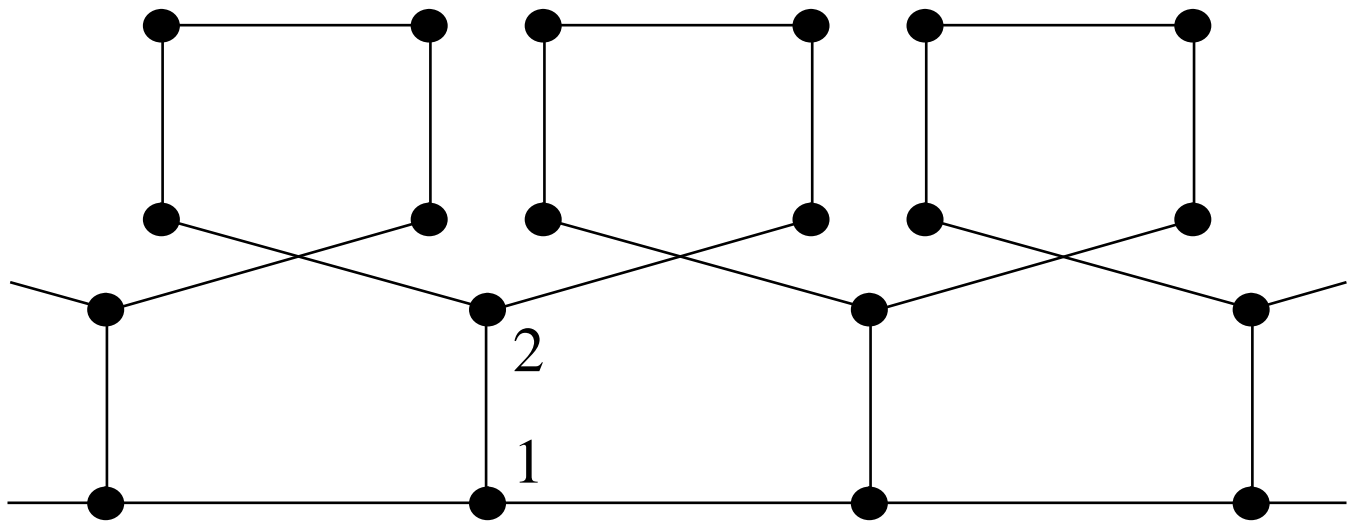
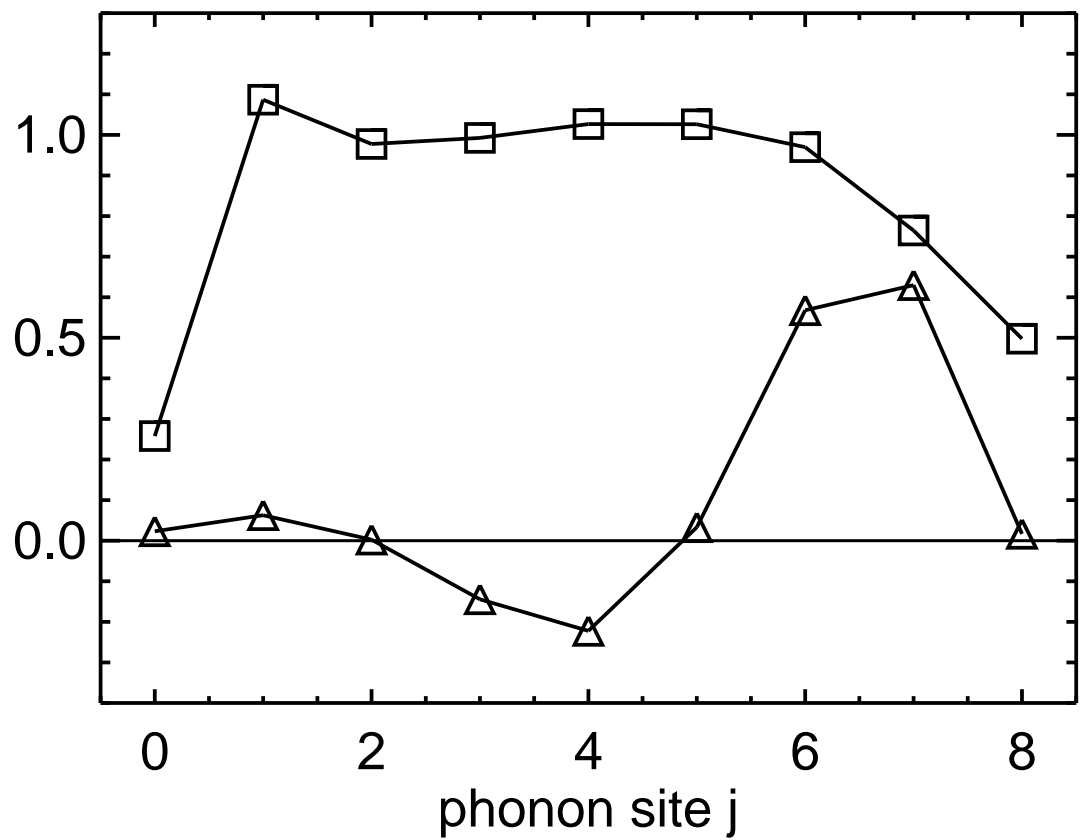
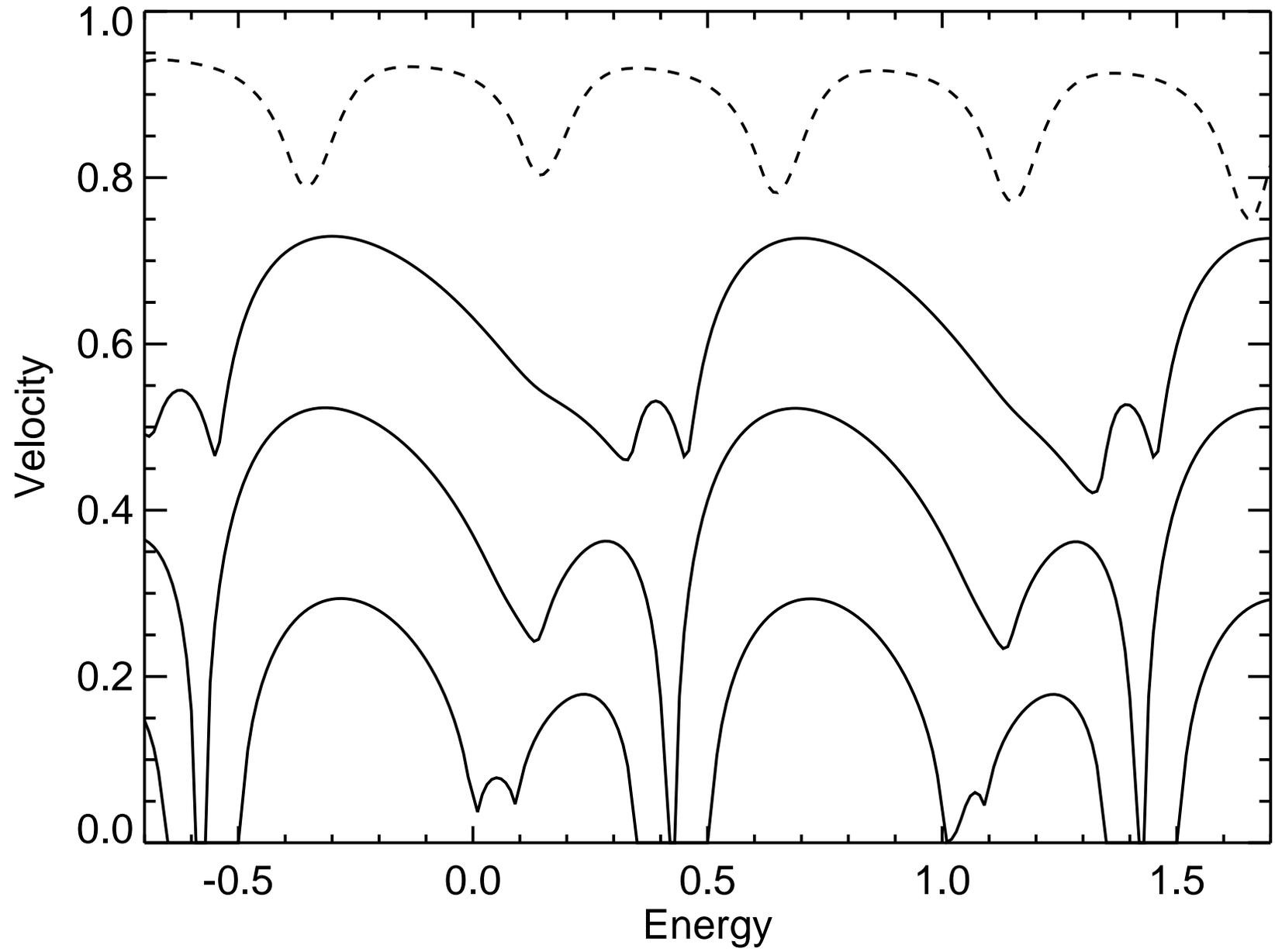
\end{document}